\documentstyle[aps,12pt,preprint,epsfig]{revtex}

\def \to {\rightarrow}
\def \beq {\begin{equation}}
\def \eeq {\end{equation}}
\def \ba {\begin{eqnarray}}
\def \ea {\end{eqnarray}}
\def \jpsi {J/\psi}

\def \< {\left <}
\def \> {\right >}

\begin{document}
\preprint{PKU-TP-97-23}
\draft

\title{Diffractive $\jpsi$ production as a probe of the gluon component
        in the Pomeron}

\author{Feng Yuan}
\address{\small {\it Department of Physics, Peking University, Beijing 100871, People's Republic
of China}}
\author{Kuang-Ta Chao}
\address{\small {\it China Center of Advanced Science and Technology (World Laboratory), Beijing 100080,
        People's Republic of China\\
      and Department of Physics, Peking University, Beijing 100871, People's Republic of China}}

\maketitle
\begin{abstract}

Presented here is a study of the large $p_T$ $\jpsi$ production in hard diffractive
process by the pomeron exchange at the Fermilab Tevatron.
We find that this process $(p\bar p \to p+\jpsi+X)$ can be used to probe the
gluon content of the pomeron and to measure the gluon fraction of
the pomeron.
And the diffractive direct $\jpsi$ production can also provide another crucial
test for the color-octet fragmentation mechanism.
Using the {\it renormalised} pomeron flux factor $D\approx 1/9$, the single
diffractive $\jpsi$ production cross section at large $p_T$ ($\geq 8GeV$)
is found to be of order of $0.01 {\rm nb}$,
and the ratio of the single diffractive to the non-diffractive $\jpsi$
production is $0.65\pm 0.15\%$ for the gluon fraction $f_g=0.7\pm 0.2$.
\end{abstract}


In the past few years there has been a renaissance of interest in diffractive
scattering. Diffractive processes in hadron collisions are well described by
the Regge theory in terms of the pomeron ($I\!\! P$) exchange\cite{pomeron,report}.
The pomeron carries quantum numbers of the vacuum, so it is a colorless entity
in QCD language, which may lead to the ``rapidity gap" events in experiments
However, the nature of pomeron and its reaction with hadrons remain a mystery.
In \cite{is}, the hard diffractive scattering processes have been suggested
to resolve the quark and gluon content in the pomeron.
That is to say, the pomeron has partonic structure, just as hadrons and nuclei.
Therefore, various processes may be considered to probe the partonic
structure of the pomeron at high energy colliders\cite{is,bcss,fs,bi}.

On the experimental side, the UA8 collaboration at CERN $Sp\bar p S$ collider
have studied diffractive dijet production at $\sqrt{s}=630GeV$\cite{ua8}, which
indicate a dominant hard partonic structure of the pomeron.
The H1 and ZEUS  have studied the diffractive deep inelastic scattering (DDIS)
and dijet photoproduction in $ep$ collision at $\sqrt{s}=300GeV$\cite{hera}.
From these measurements, ZEUS determined that the gluon fraction of the pomeron
$f_g$ is in the range of $0.3<f_g<0.8$, and H1 determined the quark fraction
of the pomeron $f_q\approx 0.2$.
The partonic structure of the pomeron is also studied recently by the
CDF collaboration at the Fermilab Tevatron through the diffractive $W$ production
\cite{dw} and dijet production\cite{dijet}, which give further evidence for the hard
partonic structure of the pomeron.
The combination of these two measurements determined the gluon
fraction of the pomeron to be $0.7\pm 0.4$.

In this paper, we will discuss another diffractive process, the single
diffractive (SD) $\jpsi$ production at large $p_T$, (shown in Fig.1),
\beq
p+\bar p \to p+\jpsi+X.
\eeq
By the following calculations, we will show that the SD $\jpsi$ production is sensitive
to the gluon fraction of the pomeron. So, the measurement of diffractive $\jpsi$
production at the Fermilab Tevatron would provide a probe of
the gluon distribution in the pomeron.
Importantly, at hadron colliders the $\jpsi$ production is of special
significance because it has extremely clean signature through its leptonic
decay modes.
Furthermore, the SD $\jpsi$ production is also interesting
to the study of heavy quarkonium production mechanism, which is another
hot topic in the past few years.

For a long time, it was believed that the heavy quarkonium production
at large $p_T$ dominantly comes from the leading order color-singlet processes\cite{schuler}.
But, as pointed out by Braaten and Yuan
\cite{frag}, the fragmentation contributions may dominate over those from
leading-order processes at sufficiently large $p_T$, although the
fragmentation processes are of higher order in strong coupling constant
$\alpha _s$.
However, the measurements of large $p_T$ charmonia production
from the CDF at the Tevatron show a large excess of direct production
(excluding the contribution from $b$ decays and the feeddown from $\chi _c$)
both for $\jpsi$ and $\psi ^{\prime }$\cite{cdfbdecay}\cite{cdf3}.
The experimental measurement is a factor of $30\sim 50$ larger than the theoretical
prediction of the Color-Singlet Model even if including the fragmentation
contributions.
Motivated by this ``surplus'' problem, a new mechanism for heavy quarkonium
production at large $p_T$ in hadronic collisions, named as Color-Octet gluon
fragmentation has been proposed\cite{com}, which is based on the
factorisation formalism of non-relativistic quantum chromodynamics (NRQCD)\cite{nrqcd}.
In the past few years, applications of the NRQCD factorisation formalism
to $\jpsi$($\psi ^{\prime }$) production at various experimental facilities
have been studied \cite{annrev}.

According to NRQCD factorisation formalism, the
gluon fragmentation to $\jpsi$ production can be factorized as
\ba
\label{expansion} D_{g\to \jpsi}(z,\mu ^2)=\sum\limits_n d_{g\to n}(z,\mu
^2)\langle {\cal O}_n^{\jpsi}\rangle ,
\ea
where $z$ is the longitudinal momentum fraction carried by the produced $\jpsi$
in gluon fragmentation, $\mu =2m_c$ is the fragmentation scale.
$d_{g\to n}$ represent the short-distance coefficients associated with the
perturbative subprocesses in which a $c\bar c$ pair is produced in a
configuration denoted by $n$ (angular momentum $^{2S+1}L_J$ and color index
1 or 8). $\langle {\cal O}_n^{\jpsi}\rangle $ are the long distance
nonperturbative matrix elements demonstrating the probability of a $c\bar c$
pair evolving into the physical state $\jpsi$. The short-distance
coefficients $d_{g\to n}$ can be obtained from perturbative calculations in
powers of coupling constant $\alpha _s$. $\langle {\cal O}_n^{\jpsi}\rangle $
consist of two kinds of matrix elements, {\it i.e.}, the color-singlet and
color-octet matrix elements (according to the color index is 1 or 8).

For $\jpsi$ production in gluon fragmentation, the color-octet matrix
element $\langle {\cal O}_8^{\jpsi}({}^3S_1)\rangle $ is smaller than the
color-singlet matrix element $\langle {\cal O}_1^{\jpsi}({}^3S_1)\rangle $
by a factor of order $v^4$ according to the NRQCD velocity scaling rules.
However, the short-distance coefficient for the color-octet term in Eq.(\ref
{expansion}) is larger than that for the color-singlet term by a factor of
order $1/\alpha _s^2$. Numerical results show that color-octet contributions
are 50 times larger than color-singlet contributions\cite{com}. In the
following calculations, we neglect the color-singlet term in gluon
fragmentation in Eq.(\ref{expansion}), and only consider the color-octet
gluon fragmentation. The leading-order color-octet gluon fragmentation to $%
\jpsi$ production gives \cite{com}
\beq
\label{frag}
D_{g\to \jpsi}^{(8)}(z,\mu ^2)=\frac{\pi \alpha _s(2m_c)}{24}%
\frac{\langle {\cal O}_8^{\jpsi}({}^3S_1)\rangle }{m_c^3}\delta (1-z).
\eeq

In our calculations, the effects of the evolution of gluon fragmentation
function with scale $\mu ^2$ are neglected, which may introduce some error.
However, as argued in \cite{sp}, including evolution would not necessarily
be an improvement, since naive Altarelli-Parisi equations do not respect the
phase-space constraint $D_{g\to \jpsi}(z,\mu ^2)=0$ for $z<m_{\jpsi}^2/\mu^2 $
\cite{ap}.

As shown in Fig.1 and Fig.2, the SD process $p\bar p\to p+\jpsi+X$ by pomeron
exchange consists of three steps.
First, a pomeron is emitted from the proton with a small squared
momentum transfer, $t=(p_i-p_f)^2$, where $p_i$ and $p_f$ are the momenta
of the initial and the final states of the proton respectively.
Second, partons interaction between the pomeron and the antiproton
takes place in the large momentum transfer processes (see Fig.2).
In the third step, $\jpsi$ is produced via the fragmentation processes.
Because the fragmentation contributions dominantly come from the color-octet
gluon fragmentation process, in our calculations we calculate the 
$gg(q\bar q)\to gg$, $q(\bar q)g\to q(\bar q)g$ processes
and followed by the gluon fragmentation Eq.(\ref{frag}).

Using the pomeron factorisation formalism\cite{is,bcss}, we write the SD
$\jpsi$ production cross section as
\beq
\frac{d\sigma(p\bar p\to p+\jpsi+X)}{d\xi d t}=
        f_{I\!\! P/p}(\xi,t)d\sigma(I\!\! P\bar p\to \jpsi+X),
\eeq
where $\xi$ is the momentum fraction of the proton carried by the pomeron.
$f_{I\!\! P/p}$ is the pomeron ``flux" factor,
\ba
\nonumber
f_{I\!\! P/p}(\xi,t)&=&\frac{d^2\sigma_{SD}/d\xi dt}{\sigma_T^{I\!\! Pp}(s^\prime,t)}
        =\frac{\beta_1^2(t)}{16\pi}\xi^{1-2\alpha(t)}F^2(t)\\
        &=&K\xi^{1-2\alpha(t)}F^2(t).
\ea
Following \cite{ren}, the parameters are choosen as
\beq
K=0.73GeV^{-2},~~\alpha(t)=1+0.115+0.26(GeV^{-2})t,~~F^2(t)=e^{4.6t}.
\eeq

In our calculations, we use the {\it renormalized} flux factor $D\cdot f(\xi,t)$
\cite{ren}, which may preserve the shapes of $M^2$ and $t$ distributions in 
SD and predict the experimentally observed SD cross section at all energies.
Here $D$ is defined as
\beq
\label{factor}
D=Min(1,1/{\cal N}),
\eeq
with
\beq
{\cal N}=\int\limits_{\xi_{min}}^{\xi_{max}}d\xi\int\limits_{t=0}^\infty f_{I\!\! P/p}(\xi,t),
\eeq
where, $\xi_{min}=M_0^2/s$ with $M_0^2=1.5GeV^2$ (effective threshold) and
$\xi_{max}=0.1$ (coherence limit).
For the SD process at the Tevatron ($\sqrt{s}=1800GeV$), the
renormalized factor $D\approx 1/9$.
As a conservative estimate, we will take this as a tentative value for
$D$ in the following calculations.
(However, we should keep in mind that this flux factor $D$ has not been
well determined experimentally. If the precise value can be obtained in the
future, our results will change accordingly.)

On the parton structure functions of the pomeron, we assume the hard
form\cite{bi,dw,dijet},
\beq
\label{form}
\beta G(\beta) =(f_q+f_g) [6\beta (1-\beta)],
\eeq
where $\beta$ is the momentum fraction of the pomeron carried by the quarks and
gluons. $f_q$ and $f_g$ are the quark and gluon  fractions of the pomeron
respectively. The momentum sum rule constrains $f_q+f_g=1$.
We neglect any $Q^2$ evolution in the above parton densities of the pomeron\cite{bi}.

With these partons densities in the pomeron and the parton distribution functions
in the antiproton, $d\sigma(I\!\! P\bar p\to \jpsi+X)$ can be calculated
by employing the usual way in the parton model calculations in hadronic
collisions.
We use the MRS(A) parton distribution functions\cite{mrs} to generate the production
cross section, and set the renomalization scale and the factorization scale
both equal to the transverse momentum of the fragmenting gluon $\mu
=p_T(g)\approx p_T(\psi )$.
In gluon fragmentation, the input parameters are taken to be
\beq m_c=1.5GeV,~~\alpha _s(2m_c)=0.26,~~\langle {\cal O}_8^{
        \jpsi}({}^3S_1)\rangle =0.0106 GeV^3.
\eeq
The value of the color-octet matrix element $\langle {\cal O}_8^{\jpsi
}({}^3S_1)\rangle$ follows the fitted value in\cite{benek} by comparing the
theoretical prediction to the experimental data at the Tevatron.
In the total cross section, we also include the contributions from
the $\chi _c$ and $\psi ^{\prime }$ feeddowns through gluon fragmentation
$g\to \chi _c$ and $g\to \psi ^{\prime }$ followed by $\chi _c\to
\psi \gamma $ and $\psi ^{\prime }\to \psi X$.
The feeddown contributions give the same $p_T $ distribution of $\jpsi$ and
contribute about one third of the total prompt $\jpsi$ production cross section
(see Ref.\cite{cdf3}).
The leptonic decay branching ratio $Br(\jpsi\to \mu^+\mu^-)=0.0597$ is also
multiplied in the cross section.
The above described procedure (including the fragmentation approximation)
can reproduce the large $p_T$ $\jpsi$ production in the central region
(i.e, $|\eta(\jpsi)|<0.6$) at the Tevatron\cite{cdf3,com}.

Because the $q\bar q$ annihilation process only contributes a small portion to the
total cross section of large $p_T$ $\jpsi$ production,
the SD $\jpsi$ production is insensitive to the quark flavor number of the
pomeron. In our calculations we only consider two-quark flavors.
A pseudorapidity cut of $2.0<\eta (\psi)<4.0$ was
also performed on the produced $J/\psi$.
The diffractive variable $\xi$ and $t$ are integrated over the range of
$0<\xi<0.1$ and $|t|<1GeV^2$.
In Fig.3, We show the cross section of SD $\jpsi$ production as a
function of the minimum transfer momentum of the produced $\jpsi$.
Because the $gg$ process is the dominant process in the production
of $\jpsi$, the cross section is sensitive to the gluon component in
the pomeron (i.e., the gluon fraction $f_g$).
In this figure, we plot three curves correspondence to three different
values of the gluon fraction $f_g$.
The solid line represents the cross section $\sigma(p\bar p\to p+\jpsi+X)\times
Br(\jpsi\to \mu^+\mu^-)$ for the gluon fraction set to be $f_g=0.7$ (which is
determined by the experiments at the Tevatron\cite{dijet}).
The dotted line is for $f_g=0$ ($f_q=1$) and the dashed line for $f_g=1$
($f_q=0$).
These curves show that the SD $\jpsi$ production cross section may reach the
level of order of $0.01{\rm nb}$ (for $p_T({\rm min})=8GeV$), and therefore is observable
at the Fermilab Tevatron at present.

In the above calculations, we use the widely used parametrization of the 
gluon distribution in the pomeron, i.e., the hard form Eq.(\ref{form}).
However, the precise form of the gluon density of the pomeron is unknown at present
and this will affect the $p_T$ distribution of the SD $\jpsi$ production.
Different parametrizations will give rise to different spectra.
If the gluon in the pomeron is soft, e.g., the gluon distribution in the
pomeron behaves like,
\beq
\beta G(\beta) =6(1-\beta)^5,
\eeq
the spectra would be different.
In Fig.4, we show the $p_T$ distributions of the cross section of the SD $\jpsi$
production in both the hard gluon and soft gluon cases.
The result shown in Fig.4 is consistent with the expectation that the softer
gluon will favor $\jpsi$ production with smaller $p_T$, while the harder
gluon will favor larger $p_T$.
The differential cross sections at large $p_T$ for these
two cases are different,
but their differences are not so critical to distinguish between them.
So, in the following discussion, we mainly limit us to the hard gluon 
parametrization of the pomeron, but we will also mention the result for the
soft gluon parameterization.

As a comparison, we also calculate the non-diffractive (ND) forward $\jpsi$ production
($p+\bar p\to \jpsi +X$)
in the same kinematic region, i.e., $2.0<\eta (\psi)<4.0$.
The forward region $\jpsi$ production is also interesting to the study of
the $\jpsi$ production mechanism, because the relative contributions of different
mechanisms may vary with $\jpsi$ rapidity.
Furthermore, the comparison of forward and central region $\jpsi$ production
can provide a consistent test of the ``surplus" problem of $\psi^\prime$
and $\jpsi$ found at the Tevatron\cite{cdf3}.
In Fig.5, we plot the ND forward $\jpsi$ production cross section as a
function of $p_T({\rm min})$.
This theoretical prediction can be used to compare with the experimental
data, and may provide a further test for the color-octet gluon fragmentation
production mechanism. (The D0 collaboration at the Fermilab Tevatron have
reported the forward $\jpsi$ production data\cite{d01}. However, these data
do not exclude the contributions from $b$ decays. We hope that the prompt
$\jpsi$ production data in the forward region may be obtained in the near
future.)

In Fig.6, we plot the ratio $R(\psi)=\sigma_{SD}/\sigma_{ND}$ as a function of
$f_g$ (solid line), where $\sigma_{SD}$ is the cross section for SD $\jpsi$ production, and
$\sigma_{ND}$ is for the ND $\jpsi$ production in the same kinematic
region.
The kinematic constraints for the produced $\jpsi$ are the same for these two
processes, i.e., $p_T({\rm min})=8GeV$ and $2.0<\eta(\psi)<4.0$.
The ratio $R(\psi)$ increase from $0.1\%$ as $f_g=0$ to $0.9\%$ as $f_g=1.0$.
For $f_g=0.7\pm 0.2$, the ratio $R(\psi)$ will be $0.65\pm0.15\%$.
We must note that $R(\psi)$ is independent of the choice of the color-octet
matrix element $\langle {\cal O}_8^{\jpsi}({}^3S_1)\rangle$ because its
dependence is cancelled in the ratio $\sigma_{SD}/\sigma_{ND}$.
So, measuring this ratio $R(\psi)$ can determine the gluon fraction
$f_g$ precisely, provided that the color-octet gluon fragmentation is the
dominant mechanism for the $\jpsi$ production at large $p_T$.

One more thing that must be noted in the above calculations of the ratio
$R(\psi)=\sigma_{SD}/\sigma_{ND}$ is the approximation of neglecting
the fragmentation function smearing.
If the $p_T$ distributions of the SD and ND $\jpsi$ production are much
different, the smearing effects will influence the ration $R(\psi)$ and
the extraction of $f_g$ from this ratio.
To see these effects, we calculate the ratio $R(\psi)=\sigma_{SD}/\sigma_{ND}$
for different values of $p_T(min)$, which is shown in Table I.
From this table, we can see that the ratio $R(\psi)$ is almost a constant
(with a fluctuation less than $10\%$)
as a function of $p_T(min)$ for both the hard quark and hard gluon
cases. This implies that the ratio $R(\psi)$ is not sensitive to the smearing,
and we may neglect the smearing effects in the calculations of the
ratio $R(\psi)$ and the extraction of the gluon fraction $f_g$.

In Table I, we also give the result for the soft gluon parametrization.
The ratio $R(\psi)$ for the soft gluon is larger than the hard gluon by
a factor $\le 2$ at $p_T(min)\ge 10GeV$.

Experimentally, the non-diffractive background to the diffractive $\jpsi$
production must be dropped out to obtain useful information of the
above calculations.
Theoretically, the SD $\jpsi$ production events can be distinguished from
those non-diffractive events by performing the Rapid Gap (RG) analysis.
However, the acceptance of the RG will affect this analysis.
Here, we adopt the existing results of the background estimate obtained
by the CDF diffractive Dijet experiment, where they give the non-diffractive
background to the SD evens to be $20\%$\cite{dijet}.
By the same reason, we expect that the non-diffractive background to the
SD $\jpsi$ production is about $20\%$.

Finally, we discuss the theoretical uncertainty coming from the choice of the
factor $D$ of Eq.(\ref{factor}).
The factor $D$ represents the momentum fraction of the pomeron
carried by the hard partons with the standard pomeron flux.
In our calculations, we use the {\it renormalized} factor, which is about
$1/9$ at the Tevatron energy region.
But the value cited here is not unique, because it may change with different
choices of the parameters such as $M_0$ and $\xi_{max}$ in Eq.(\ref{factor}).
The momentum fraction $D$ can be measured in the diffractive processes
at various collider faculties.
At the Tevatron, the CDF have determined the fraction $D$ to be
$0.18\pm 0.04$\cite{dijet}, which is well below the range $0.4<D<1.6$ reported
by ZEUS.
If we adopt the CDF measurement, there must be difference
in the above calculations of the ratio $R$, which is also shown in Fig.6.
The shaded region in Fig.6 represent the range of the ratio $R$ calculated
as a function of $f_g$ by using the fraction $D=0.18\pm 0.04$. 

As discussed in previous studies\cite{com,benek}, color-octet mechanism is crucially important
to direct $\jpsi$ production (excluding the contributions from $b$ decays and
the $\chi_c$ and $\psi^\prime$ feeddowns) at large $p_T$, here the color-octet
mechanism is also crucially important to the direct $\jpsi$ production in the
diffraction region.
If only considering the color-singlet contributions (mainly coming from gluon
fragmentation) the SD direct $\jpsi$ production rate will be smaller than
the curves shown in Fig.3 by a factor of 50.
This will make the measurement of the SD direct $\jpsi$ production very difficult
at present luminosity at the Tevatron.
That is to say, the SD direct $\jpsi$ production can also be regarded as another
important test for the color-octet mechanism.

As a final remark, we note that our proposal, the diffractive $\jpsi$ production,
may also be used to extract the gluon fraction in photoproduction
at the HERA.
There are more diffractive events at the HERA than that at the Tevatron.
So, more interesting results may be obtained.
The work along this way is in progress\cite{future}.

In conclusion, in this paper we show that the diffractive $\jpsi $ production
at large $p_T$ is sensitive to the gluon fraction of the pomeron $f_g$.
The measurement of this process at the Tevatron would provide a 
determination of $f_g$.
We have also discussed the uncertainties caused by the renormalised factor
$D$ and the gluon parameterizations of the pomeron.
These uncertainties, however, can be reduced by combining other experimental
measurements such as dijet diffractive production and $W$ diffractive
production.
We believe that with the proposed SD $\jpsi$ production presented in this
paper, we will get a better understanding for the property of the pomeron.
And also, the SD direct $\jpsi$ production will provide another crucial
test for the color-octet production mechanism.

\vskip 1cm

\begin{center}
{\bf {\large {Acknowledgments}\ }}
\end{center}

We would like to thank Prof. H.A. Peng for his reading of the manuscript
and his comments.
We also thank Dr. J.S. Xu for interesting discussions,
and especially we thank Prof. H.Y. Zhou for providing us with the computer
program and many enthusiastic discussions. 
This work was supported in part by the National Natural Science Foundation
of China, the State Education Commission of China, and the State
Commission of Science and Technology of China.


\begin{table}[tbp] \centering
\begin{tabular}{|l|l|l|l|l|l|l|l|l|l|}
\hline
$P_T(\min )(GeV)$ & 5.0 & 6.0 & 7.0 & 8.0 & 9.0 & 10.0 & 11.0 & 12.0 & 13.0
\\ \hline
$R(\psi )$ for hard quark $(\%)$ & 0.17 & 0.16 & 0.15 & 0.14 & 0.14 & 0.14 & 
0.14 & 0.14 & 0.14 \\ \hline
$R(\psi )$ for hard gluon $(\%)$ & 0.98 & 0.93 & 0.90 & 0.86 & 0.84 & 0.85 & 
0.86 & 0.88 & 0.87 \\ \hline
\end{tabular}
\caption{The ratio $R(\psi)=\sigma_{SD}/\sigma_{ND}$ as a function of $p_T(min)$. \label{key}}
\end{table}

\begin{figure}
\begin{center}
\epsfig{file=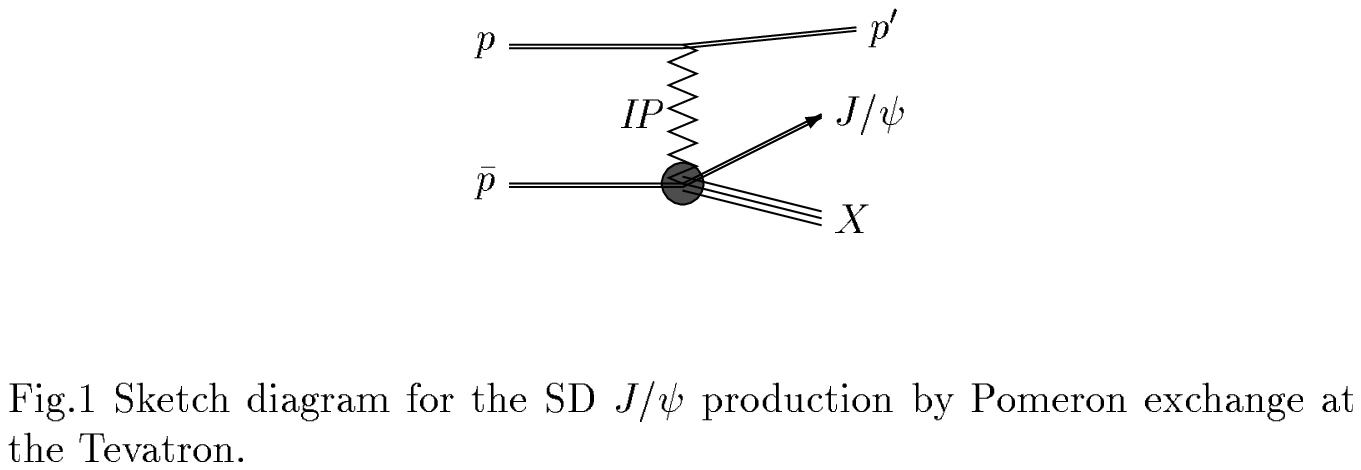}
\end{center}
\end{figure}
\begin{figure}
\begin{center}
\epsfig{file=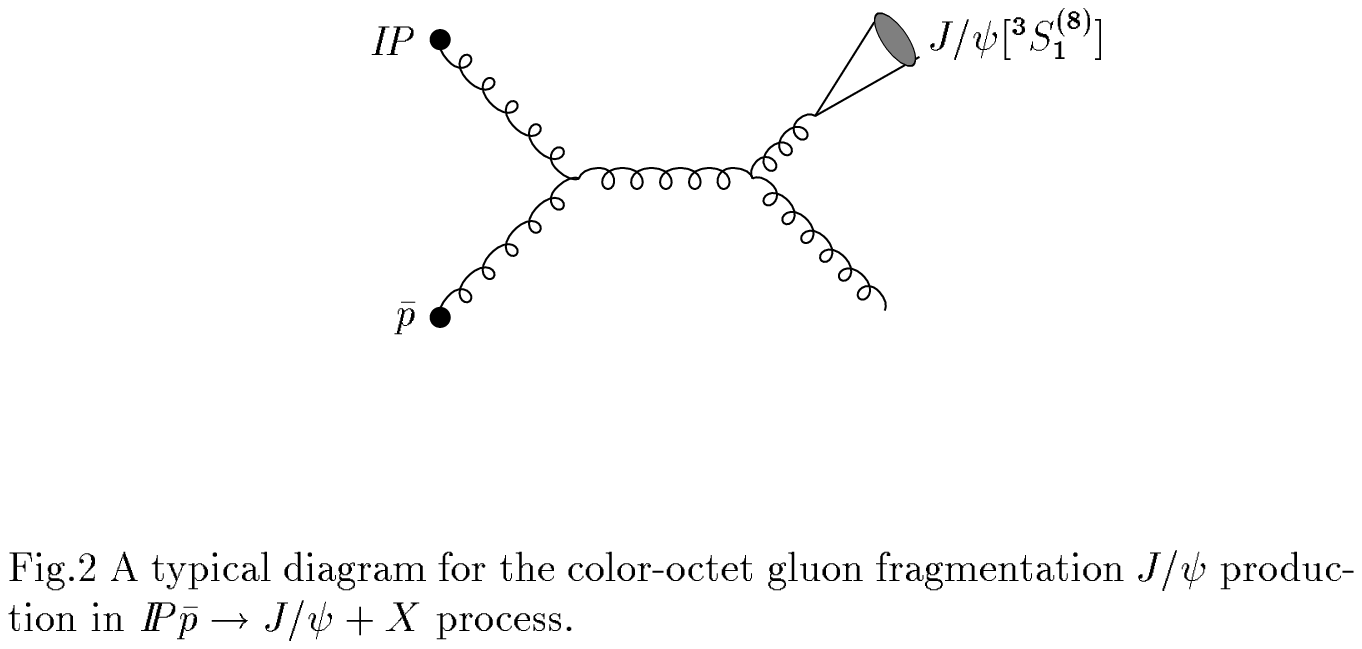}
\end{center}
\end{figure}
\begin{figure}
\begin{center}
\epsfig{file=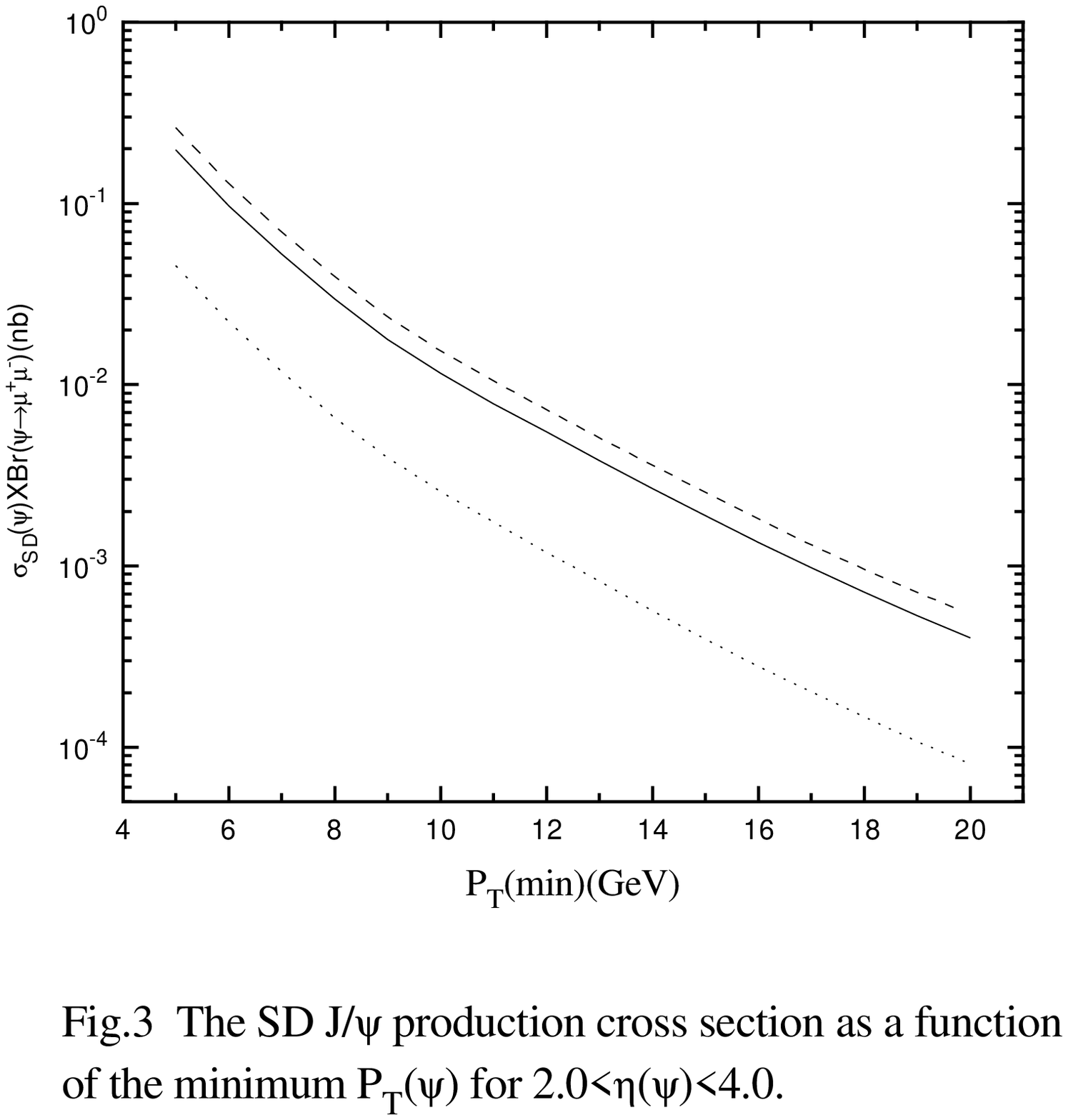}
\end{center}
\end{figure}
\begin{figure}
\begin{center}
\epsfig{file=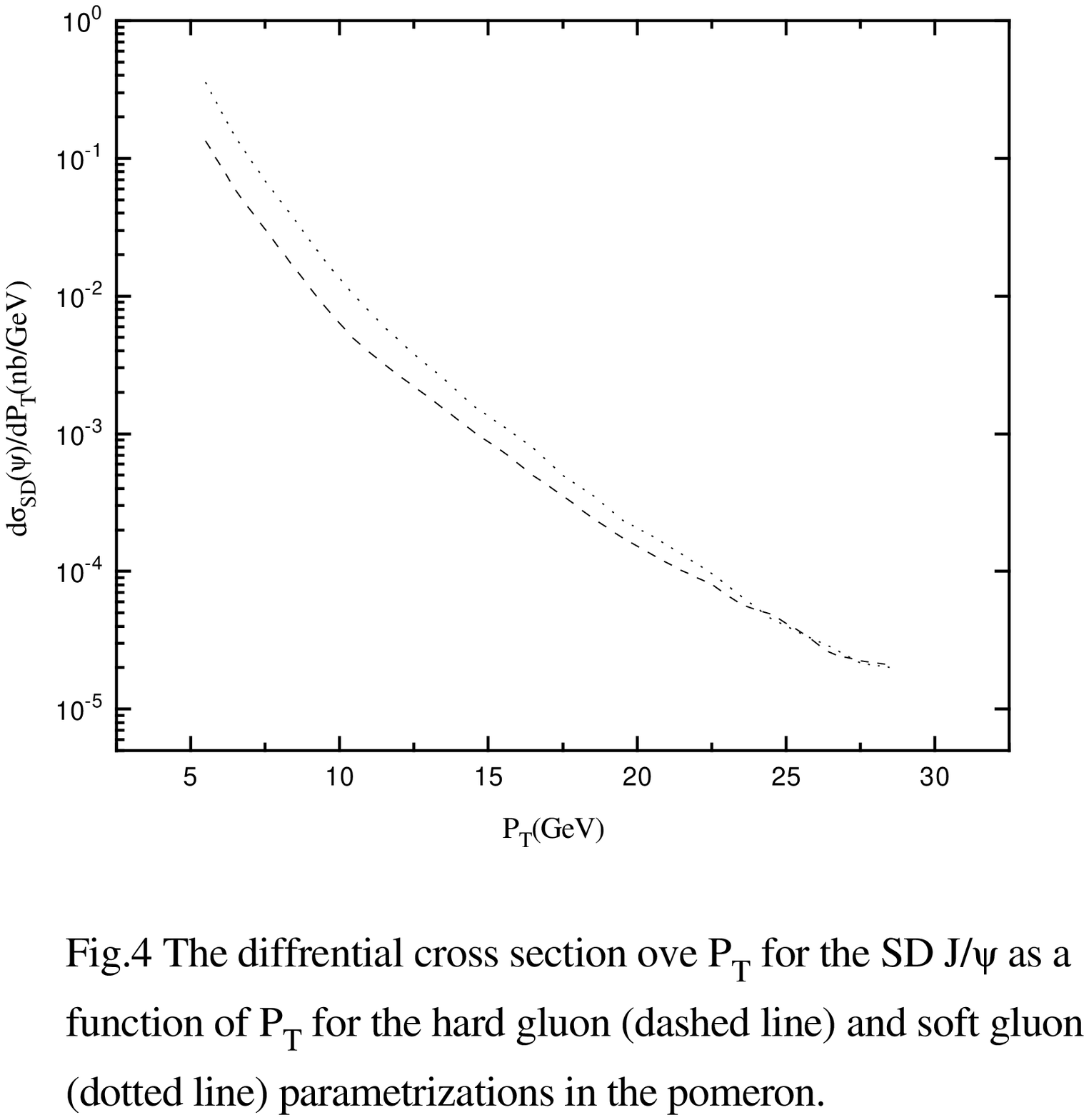}
\end{center}
\end{figure}
\begin{figure}
\begin{center}
\epsfig{file=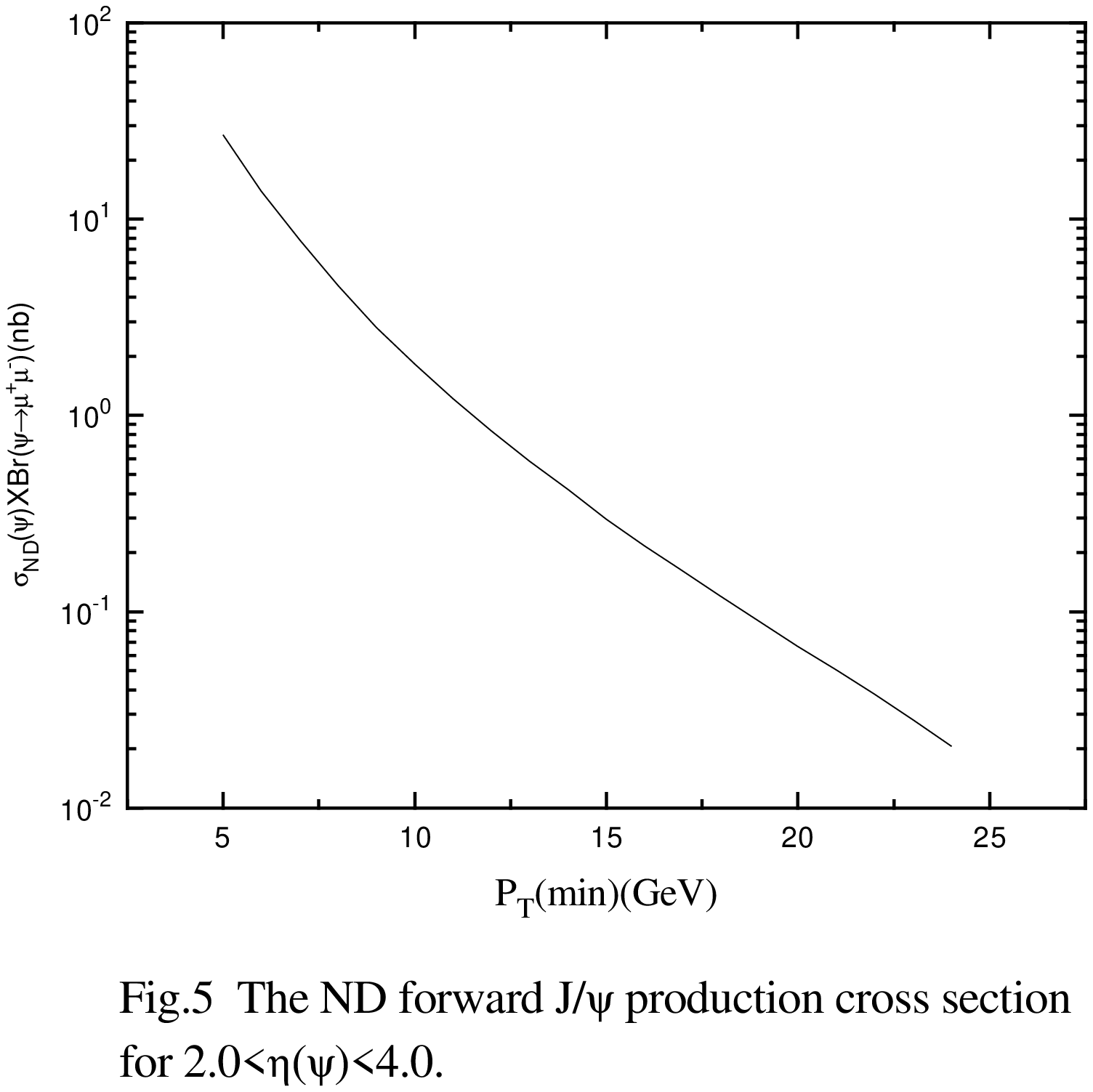}
\end{center}
\end{figure}
\begin{figure}
\begin{center}
\epsfig{file=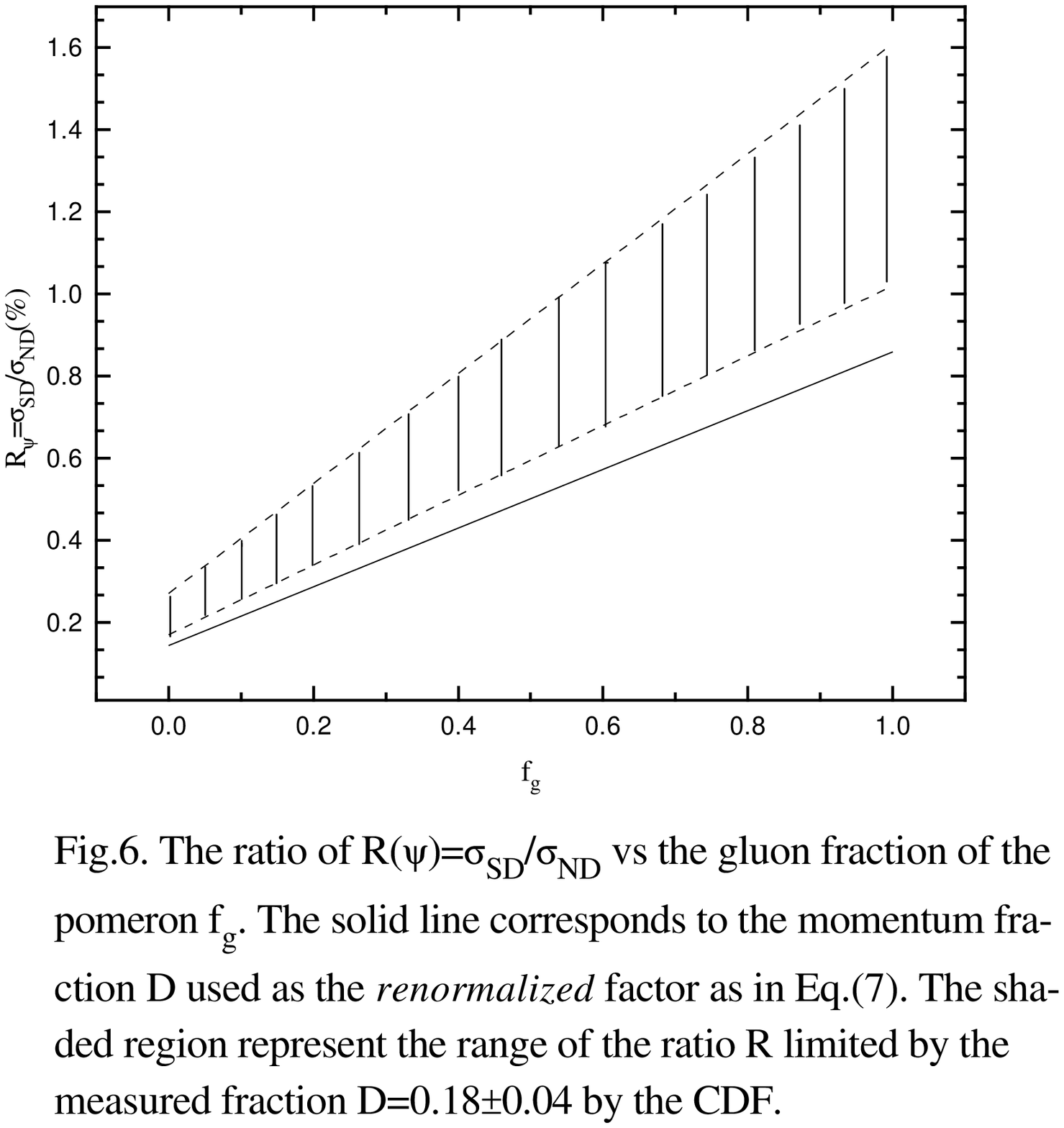}
\end{center}
\end{figure}

\end{document}